\documentclass[%
reprint,groupedaddress,
amsmath,amssymb,aps,prl]{revtex4-2}

\usepackage{graphicx}
\usepackage{dcolumn}
\usepackage{bm}
\usepackage{soul,color}
\usepackage{float}
\usepackage{comment}
\usepackage{lineno}
\usepackage{blindtext}

\newcommand{\emailx}[1]{\move@AF\def\@affil{{\normalfont\,#1\strut}{}}}%
\newcommand{\p}{\partial}

\newcommand{\vta}{\vartheta}
\newcommand{\om}{\omega}

\newcommand{\nn}{\nonumber}
\newcommand{\ta}{\theta}

\newcommand{\wt}{\tilde}
\newcommand{\wh}{\widehat}

\newcommand{\cW}{{\cal W}}
\newcommand{\cA}{{\cal A}}

\newcommand{\be}{\begin{equation}}                                       
	\newcommand{\ee}{\end{equation}}
\newcommand{\ba}{\begin{eqnarray}}
	\newcommand{\ea}{\end{eqnarray}}
\newcommand{\bref}[1]{(\ref{#1})}

\newcommand{\bi}[1]{\bibitem{#1}}\newcommand{\lab}[1]{\label{#1}}

\newcommand{\bsub}{\begin{linenomath}\begin{subequations}}                      
		\newcommand{\esub}{\end{subequations}\end{linenomath}}

\begin{document}
	\preprint{APS/123-QED}

	\title{Hyperparametric solitons in nondegenerate optical parametric oscillators}
	\author{Haizhong Weng$^{1,2}$, Xinru Ji$^{3}$, Mugahid Ali$^{1}$, Edward H. Krock$^{1}$, Lulin Wang$^{1}$, Vikash Kumar$^{1}$, Weihua Guo$^4$, Tobias J. Kippenberg$^3$, John F. Donegan$^{1,*}$ and Dmitry V. Skryabin$^{5,6,7,**}$}
	\address{
		$^1$\mbox{School of Physics, CRANN, AMBER, and CONNECT, Trinity College Dublin, D02 PN40, Dublin 2, Ireland}\\
		$^2$\mbox{ Center for Heterogeneous Integration of Functional Materials and Devices, Yongjiang Laboratory, 315202, Ningbo, China}\\
		$^3$\mbox{Institute of Physics, Swiss Federal Institute of Technology Lausanne (EPFL), CH-1015 Lausanne, Switzerland}\\
		$^4$\mbox{Wuhan National Laboratory for Optoelectronics, and School of Optical and Electronic Information,} \mbox{Huazhong University of Science and Technology, 430074, Wuhan, China}\\
		$^5$\mbox{Department of Physics, University of Bath, Bath, BA2 7AY,  United Kingdom}\\
		$^6$\mbox{Centre for Photonics, University of Bath, Bath, BA2 7AY, United Kingdom}\\
		$^7$\mbox{National Physical Laboratory,  Teddington, TW11 0LW, United Kingdom}\\
		$^{*}$Corresponding author: jdonegan@tcd.ie\\
		$^{**}$Corresponding author: d.v.skryabin@bath.ac.uk
	}	
	
	\date{\today}
	\begin{abstract}
	Dissipative solitons and their associated low-noise, chip-scale frequency combs hold great potential for applications in optical communications, spectroscopy, precision time-keeping, and beyond. These applications drive interest in shifting soliton spectra to frequency bands far detuned from the telecom's C-band pump sources. Recent demonstrations have utilized second-harmonic generation and degenerate optical parametric oscillators (OPOs) to shift soliton combs away from the primary pump. However, these approaches lack the tunability offered by nondegenerate OPOs. This work presents a proof-of-principle demonstration of solitons in a silicon-nitride microresonator-based nondegenerate OPO system with engineered dispersion and optimized coupling rates. By pumping a relatively low-Q resonance in the C-band, we excite a signal soliton comb centred around a far-detuned, high-Q O-band resonance. This process also generates repetition-rate-locked combs at the pump and idler frequencies, with the latter occurring at a wavelength beyond 
	2$\mu$m. We demonstrate that the solitons supported by this platform are distinct from other families of dissipative solitons and call them - hyperparametric solitons. They emerge when the narrow-band signal mode, phase-matched under negative pump detuning, reaches sufficient power to drive bistability in the parametric signal. We investigate the properties of hyperparametric solitons, including their parametrically generated background and multisoliton states, both experimentally and through theoretical modelling.
	\end{abstract}
	
	\maketitle

	Microresonator optical frequency combs (microcombs) have made a profound impact on modern-day nonlinear integrated photonics and revolutionised fundamental research in optical solitons \cite{pasq,kipgaeta,did,bowers}. The unique features of such microresonator combs include their compactness and high repetition rates,  with potential for mass production. 
	At the same time, dissipative solitons represent the most application-relevant class of low-noise microcombs~\cite{kipgaeta}. Broadening and stabilising soliton spectra has produced some of the most exciting recent research directions in this area. Spectra of solitons, typically centred around the pump wavelength, can span an octave in dispersion-engineered resonators~\cite{optica1,optica2,loncar1} and may be further broadened by applying a second pump that also serves to stabilise the comb repetition rates and offset frequencies~\cite{pascal,kartik,kartik1},
	and to initiate multi-colour soliton generation~\cite{darkbright}. 
	
	The rigid attachment of the aforementioned solitons, known as dissipative Kerr solitons~\cite{kipgaeta}, to a specific pump wavelength, which also builds the soliton background, may also limit the comb spectral bandwidth. Expanding the spectral range of dissipative solitons into the mid-infrared region, as well as across other telecom bands, and into the visible and even UV parts of the spectrum, while still utilising the 1550nm (C-band) pump, will significantly enhance the impact of microcombs in telecommunications, spectroscopy, and metrology.
	
	By utilizing nonlinear frequency conversion processes, 
	e.g., Raman scattering, second and third harmonic generation, and parametric frequency conversion, it is possible to design microresonators that generate microcombs far away from the pump frequency. For example, Kerr solitons in silica glass microresonators that are pumped in the C-band have been shown to generate Stokes-pair solitons through the Raman effect, allowing tunability across the L and U communication bands of optical fibers~\cite{stokes1,stokes2}. Second-harmonic generation in thin-film lithium niobate (LN) microresonators, which exhibit a combination of strong Kerr and second-order nonlinearities, have been employed to convert C-band solitons into non-solitonic near-visible combs~\cite{vahala}
	and to demonstrate two-colour solitons~\cite{skr1}.
	
	\begin{figure*}[t]
		\includegraphics[width=\textwidth]{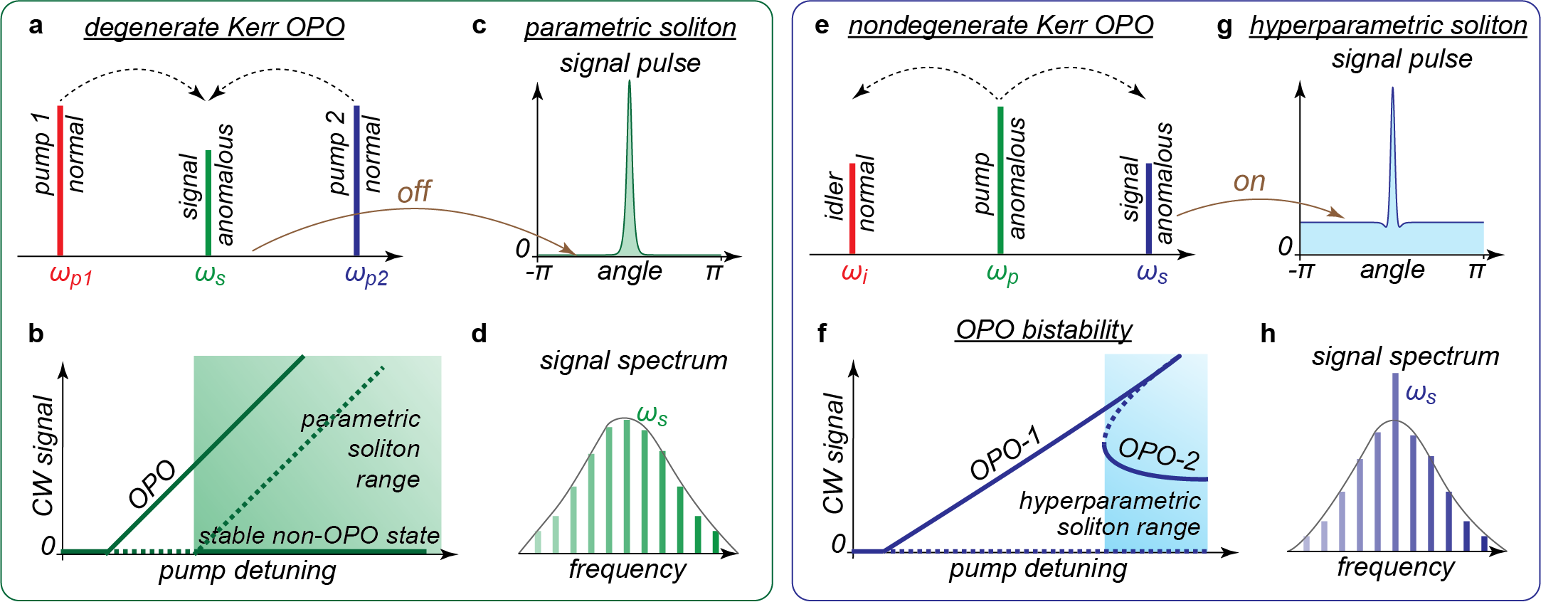}  
		\caption{
			\textbf{Parametric solitons in degenerate Kerr microresonator OPOs~\cite{moille} versus hyperparametric solitons in nondegenerate OPOs reported here.}
			{\bf a,e,} Linearly phase-matched (a) degenerate and (e) nondegenerate four-wave mixing processes.
			{\bf b,f,} (b) Coexistence of the OPO and stable non-OPO regimes in the degenerate case, and (f) bistability between two distinct OPO states in the nondegenerate case. Dashed lines indicate unstable states.
			{\bf c,d,g,h,} Parametric (c,d) and hyperparametric (g,h) solitons in the signal field and their respective spectra. 
			Text in panels (a) and (e) also specifies whether the pump, $A_{p}$, signal, $A_{s}$, and idler, $A_{i}$, components fall into regions of normal ($D_2<0$) or anomalous ($D_2>0$) dispersion.
		} 
		\label{f1}\end{figure*}

	Parametric frequency conversion is, however, of primary interest in the context of our work. By engineering phase-matched parametric frequency conversion relying on either $\chi^{(2)}$ or Kerr nonlinearity, one can create optical parametric oscillators (OPOs), also generating signals far detuned from the pump frequency
	\cite{vah0,gaeta0,moss,murdoch,fujii,perez,kipp1,kartik11,gaeta,marandi1,fejer,gaeta1,victor,tang}, see Figs. \ref{f1}a,b for a conceptual illustration of both degenerate and nondegenerate Kerr OPOs. In Kerr microresonator OPOs, the parametric gain is proportional to the coupled-in, i.e., intraresonator, pump power. As the gain increases, when the pump frequency is tuned closer to the resonance, the OPO signal can start coexisting with a stable non-OPO state, see Fig. \ref{f1}b. In the case of degenerate OPOs,  this coexistence has been a key prerequisite for recent demonstrations of parametric solitons making trains of signal pulses having a zero (non-OPO) background~\cite{bruch,engl,moille}, see Figs. \ref{f1}c,d. 
	
	In particular, parametric solitons have been demonstrated using $\chi^{(2)}$ nonlinearity for parametric down-conversion, from $775$nm pump to a $1550$nm signal, in an AlN microresonator~\cite{bruch}. A similar soliton experiment was later conducted in a fibre loop resonator, which included a short piece of $\chi^{(2)}$-active fibre for parametric gain~\cite{engl}. The reference~\cite{moille} used an all-Kerr SiN microresonator with the dispersion profile phase-matched for the degenerate four-wave mixing process, $2f_\mathrm{signal}=f_{\mathrm{pump},1}+f_{\mathrm{pump},2}$, with the pumps close to $1.0\mu$m and $1.5\mu$m creating parametric gain for the degenerate or near-degenerate signal photons at $1.2\mu$m, see Figs. \ref{f1}a-d. (Here and below $f$'s with various subscripts are used to annotate optical frequencies.)
	
	Common features of the parametric solitons observed in \cite{bruch,engl,moille} were device operation near the degeneracy point providing a zero background for soliton pulses and the coexistence of solitons with identical intensity profiles and phases differing by $\pi$~\cite{engl,moille}. The zero-background property also means that the soliton central frequency does not dominate the rest of the signal comb, see Fig. \ref{f1}d. This is unlike when a soliton is spectrally centred on the pump frequency and is governed by the classic form of the Lugiato-Lefever model~\cite{kipgaeta}. 
	
	Degenerate optical parametric oscillators (OPOs) are efficient for frequency conversion of the pump light, but they lack the tunability of nondegenerate OPOs with significant spectral separation between signal and idler, which is achieved by controlling phase matching. Indeed, in degenerate OPOs or second-harmonic generation, adjusting the phase-matching condition only controls the conversion efficiency but not the frequency of the generated signal. In recent developments, nondegenerate OPOs operating in the continuous-wave (CW) regime within microresonators and waveguides have achieved significant improvements in both tunability and efficiency
	~\cite{vah0,gaeta0,moss,murdoch,fujii,perez,kipp1,kartik11,gaeta,marandi1,fejer,gaeta1,victor,tang}. However, despite the wealth of results in the CW domain, no microresonator nondegenerate OPO that generates soliton microcombs has been demonstrated to date. 
	Our current work fills this gap by presenting  a
	new class of dissipative optical solitons, and highlighting critical differences with solitons in degenerate OPOs.

	\begin{figure*}[t]
		\includegraphics[width=\textwidth]{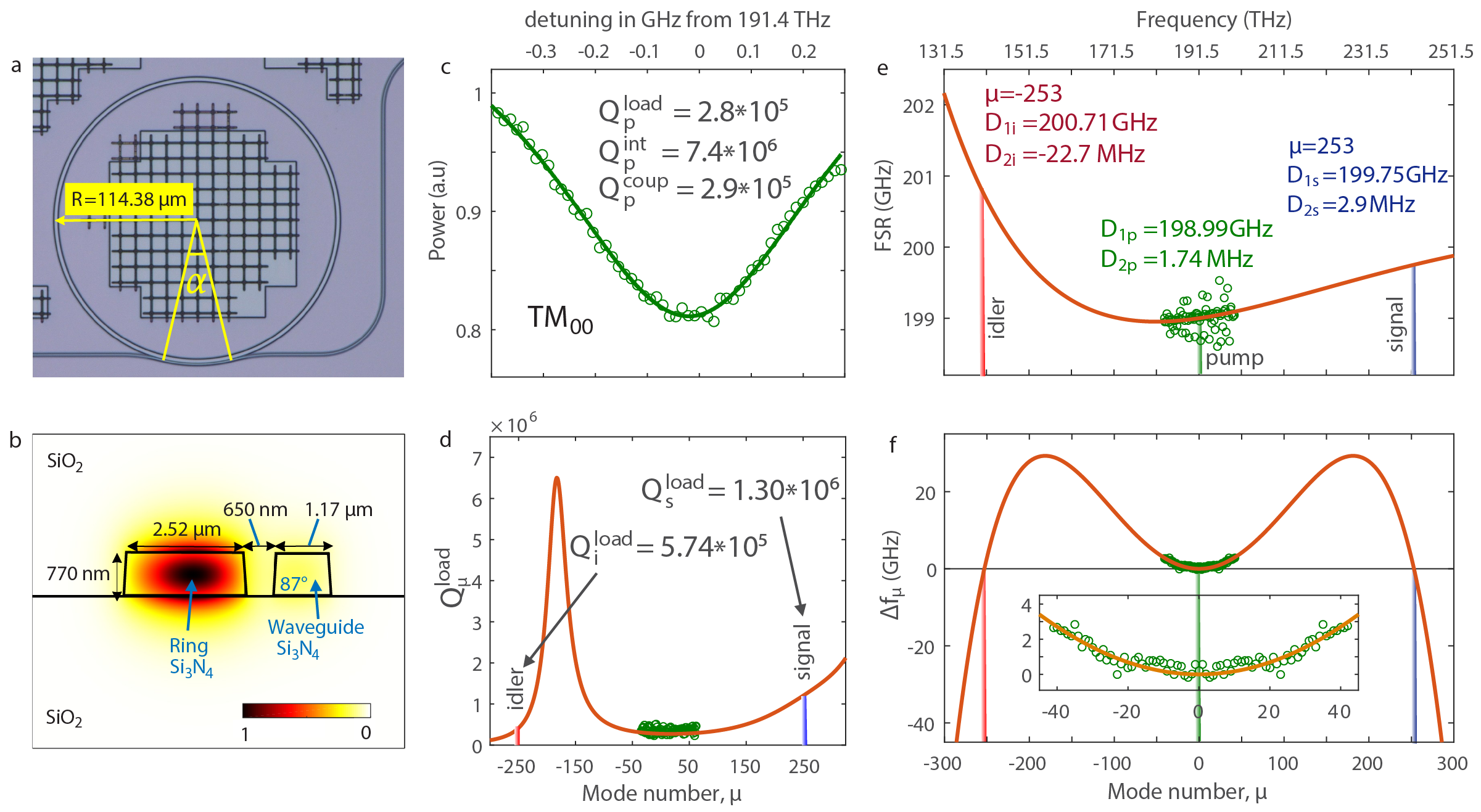}  
		\caption{{\bf Resonator geometry, loss, dispersion and phase matching profiles.}
			\textbf{a,} Microscope image of a ring resonator with a pulley coupler. 
			\textbf{b,} Dimensions of the resonator and bus waveguide.
			\textbf{c,} Measured resonance at the pump frequency, $191.5$THz. 
			$Q_p^{\mathrm{load}}$, $Q_p^{\mathrm{intr}}$ and $Q_p^{\mathrm{coup}}$ are the loaded
			(total), intrinsic and coupling quality factors, respectively. 
			Close values of $Q_p^{\mathrm{load}}$ and $Q_p^{\mathrm{coup}}$ highlight strong over-coupling conditions.
			\textbf{d,} Simulated $Q_\mu^{\mathrm{load}}$ vs mode number 
			showing steep rise above $\mu=150$ ($220$THz) and high maximum corresponding to the anti-resonance coupling condition around $\mu=-180$ ($154$THz), $\alpha=25^{\circ}$. 
			\textbf{e,} Free-spectre range (FSR) vs  relative mode number $\mu$ and frequency. Dispersion is normal for decreasing and anomalous for increasing FSR.  
			\textbf{f,} Phase-matching parameter, $\Delta f_\mu$, see Eq.~\bref{e1}.
			Circles in c-f show experimental points.} 
		\label{f2}\end{figure*}
	
	\begin{figure*}[t]
		\includegraphics[width=\textwidth]{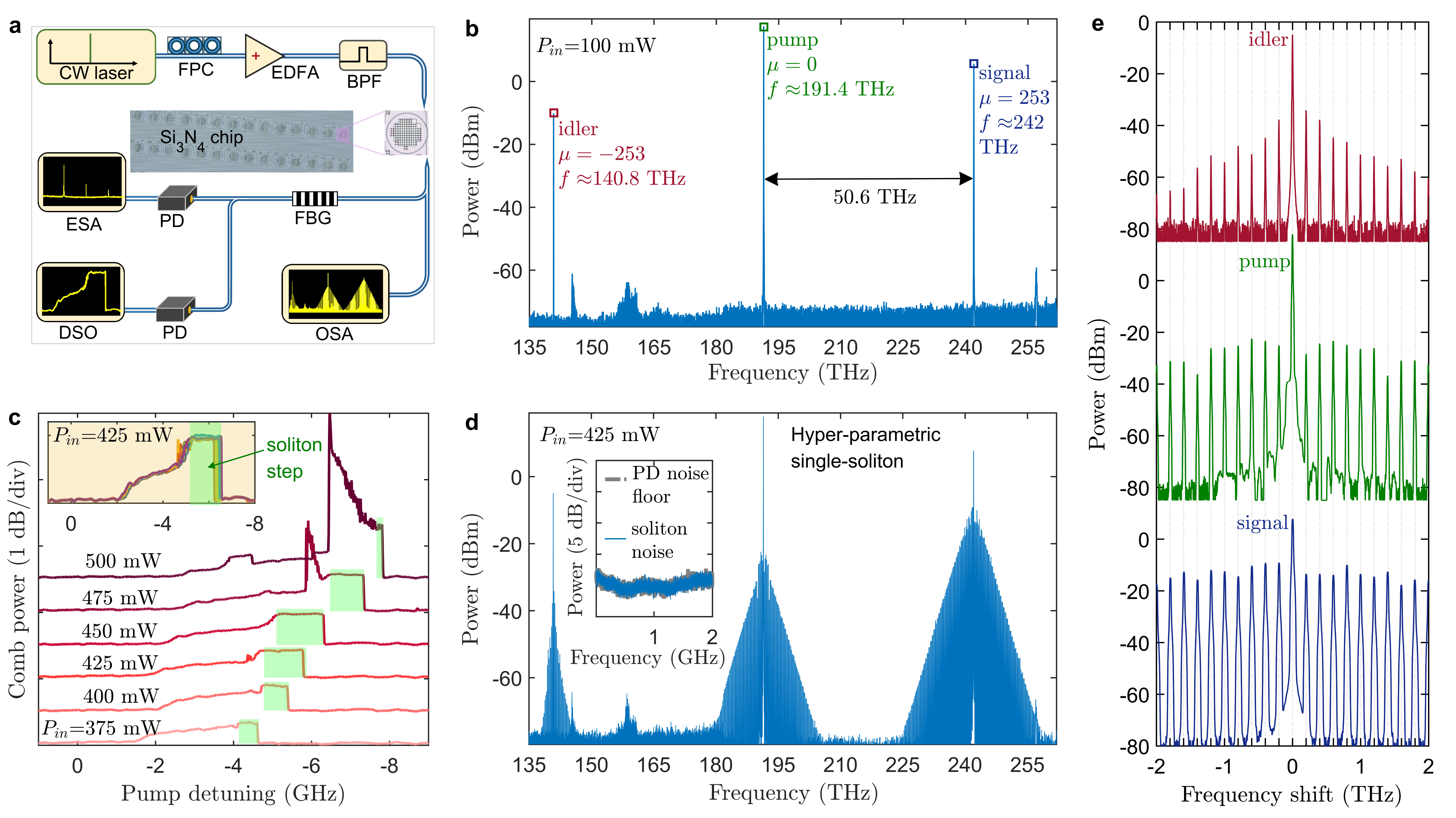}  
		\caption{{\bf Frequency comb measurement setup, soliton steps, OPO and soliton spectra.}
			\textbf{a}. Measurement setup:  FPC: polarisation controller; 
			EDFA: erbium-doped fiber amplifier; BPF: band-pass filter; FBG: fiber Bragg grating; PD: photo-diode; ESA: electrical spectrum analyser;  
			DSO: digital storage oscilloscope; OSA: optical spectrum analyser. 
			\textbf{b}. OPO spectrum measured for on-chip power of 100 mW. 
			\textbf{c}. Measured OFC power when scanning the laser across the resonance at $191.4$THz for different values ($375$, $400$, $425$, $450$, $475$ and $500$mW) of the on-chip laser power. Green shading shows the soliton steps. The inset zooms in on the $425$mW case. 
			\textbf{d}. Three-colour spectrum of a hyperparametric soliton. 
			\textbf{e}. Zoomed-in soliton spectra demonstrating that the repetition rate (line-to-line separation), $\approx 199.5$GHz, is constant across all three spectral bands and is nearest to the linear signal repetition rate $D_{1s}=199.75$GHz.}
		\label{f3}\end{figure*}

	\section*{Concept of hyperparametric solitons in nondegenerate OPO${\mathbf s}$}
	In introducing the physical concept of our work, it is natural to start by mentioning that theoretical results on bright zero-background solitons in degenerate OPOs were reported in optics as early as two decades ago~\cite{longhi,stal,trillo,firth}, and even earlier in the contexts of fluid mechanics and condensed matter physics~\cite{miles,fl,barash}.
	Theoretical investigations of bright solitons in nondegenerate OPO systems—where the signal and idler fields are well separated spectrally—were also conducted during this early period~\cite{skr,stal1}. These studies were technically and conceptually linked to the degenerate case and, therefore, found only solitons localised on a zero background, again arising in regimes where an OPO state coexists with a stable non-OPO state. Notably, we have not seen such solitons in the experiments reported below.
	
	Here, we report a class of solitons in microresonator OPOs that have not been theoretically predicted or experimentally demonstrated. We use a Si$_3$N$_4$ resonator designed and pumped (1550nm, C-band) to support a  nondegenerate phase-matched four-wave mixing process, $f_\mathrm{signal}=2f_\mathrm{pump}-f_\mathrm{idler}$, with the O-band signal near 1.25$\mu$m, and the  idler near 2$\mu$m, see Fig. \ref{f1}e. The resonator operates in the over-coupled regime, where the pulley coupler is designed to ensure that the coupling rate for the pump exceeds those for the signal and idler. We demonstrate that such conditions ensure that the signal not only increases with the pump but also depletes it sufficiently for the OPO state to fold on itself, creating a bistable loop between two distinct OPO states operating in the same resonator modes, see OPO-$1$ and OPO-$2$ in Fig. \ref{f1}f.

	Then, we experimentally and numerically demonstrate that the OPO bistability triggers the formation of signal solitons. This implies that the parametric generation does not turn off at the soliton tails since OPO-$2$ remains excited. This characteristic sets hyperparametric solitons apart from the parametric ones observed in degenerate OPOs, as illustrated in Figs. \ref{f1}c,d and \ref{f1}g,h. The term "hyperparametric solitons", which we coin here, emphasizes that parametric processes play an even more crucial role than in traditional "parametric" solitons. The latter are situated on a zero OPO background, while our solitons are based on a fully developed monochromatic parametric signal. It is worth noting that the term hyperparametric oscillations has been previously used in the context of microresonators~\cite{moss,matsko1}, but not in relation to solitons.
	
	The frequency of parametric solitons in the degenerate $\chi^{(2)}$ OPOs is restricted to half of the pump frequency~\cite{engl,bruch}. Similarly, the frequency of Kerr parametric solitons hits the mid-point between the two pumps~\cite{moille}. Thus, both setups are tunable only by adjusting the pump frequencies. In contrast, nondegenerate OPOs offer tuning capabilities in the range of several tens to up to 100 THz by choosing resonator geometries that provide suitable phase matching conditions, as has been demonstrated for CW OPOs~\cite{murdoch,fujii,perez,kipp1,kartik11,gaeta1,victor,tang}.

	\begin{figure*}[t]
		\includegraphics[width=1.\textwidth]{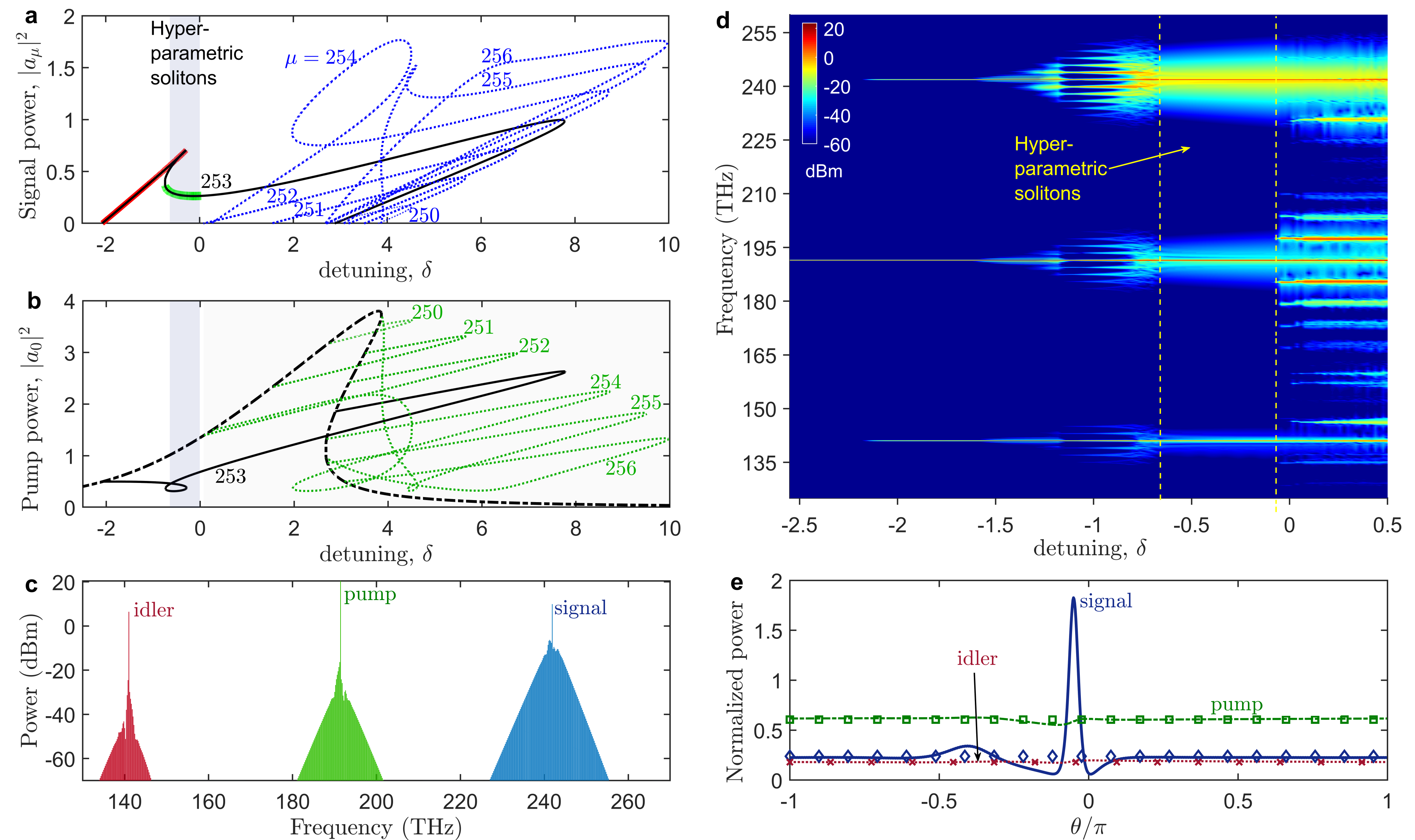}  
		\caption{
			{\bf Numerical results for parametric CW states and hyperparametric solitons.}
			\textbf{a}.~Intraresonator signal power, $|a_\mu|^2$, vs dimensionless detuning, $\delta=(f_0-f_p)/\tfrac{1}{2}\kappa_0$, for six OPO states $\mu=\pm 250,\dots,\pm 256$. $\mu=\pm 253$ OPO pair bifurcates first ($\delta\approx-2.15$) creating a bistable loop with itself; see the thick red and green lines highlighting the OPO-$1$ and OPO-$2$ states, cf. Fig.~\ref{f1}f. The shaded interval indicates the range of existence of hyperparametric solitons.
			\textbf{b}.~Dimensionless intraresonator pump power, $|a_0|^2$, corresponding to the signals shown in (a). The dashed-dotted line shows the nonlinear resonance without parametric generation, i.e., for $a_{\mu\ne 0}=0$. 
			\textbf{c}.~Three-colour output spectrum of a hyperparametric soliton computed for $\delta=-0.3215$. 
			\textbf{d}.~Numerically computed spectra for a laser frequency sweep show parametric generation of the $\mu=\pm 253$  modes at 242THz and 141THz, which is then replaced by 3-colour combs and hyperparametric solitons. 
			\textbf{e}.~Pulse profiles of the signal (bright soliton), pump (quasi-CW) and idler (quasi-CW) components of a hyperparametric soliton corresponding to the spectrum in (c). Pulse envelopes are computed as per Eq.~\bref{tp14} in Methods. Squares, diamonds, and crosses show the powers of the pump, signal, and idler components for the OPO-$2$ state making up the soliton background. On-chip power applied in the modelling is $\cW=367$mW. Scaling for the intraresonator power to use in \textbf{a,b,e} is $18.3$W, see Methods. 
		}
		\label{f4}\end{figure*}
	
	\begin{figure*}[t]
		\includegraphics[width=1.\textwidth]{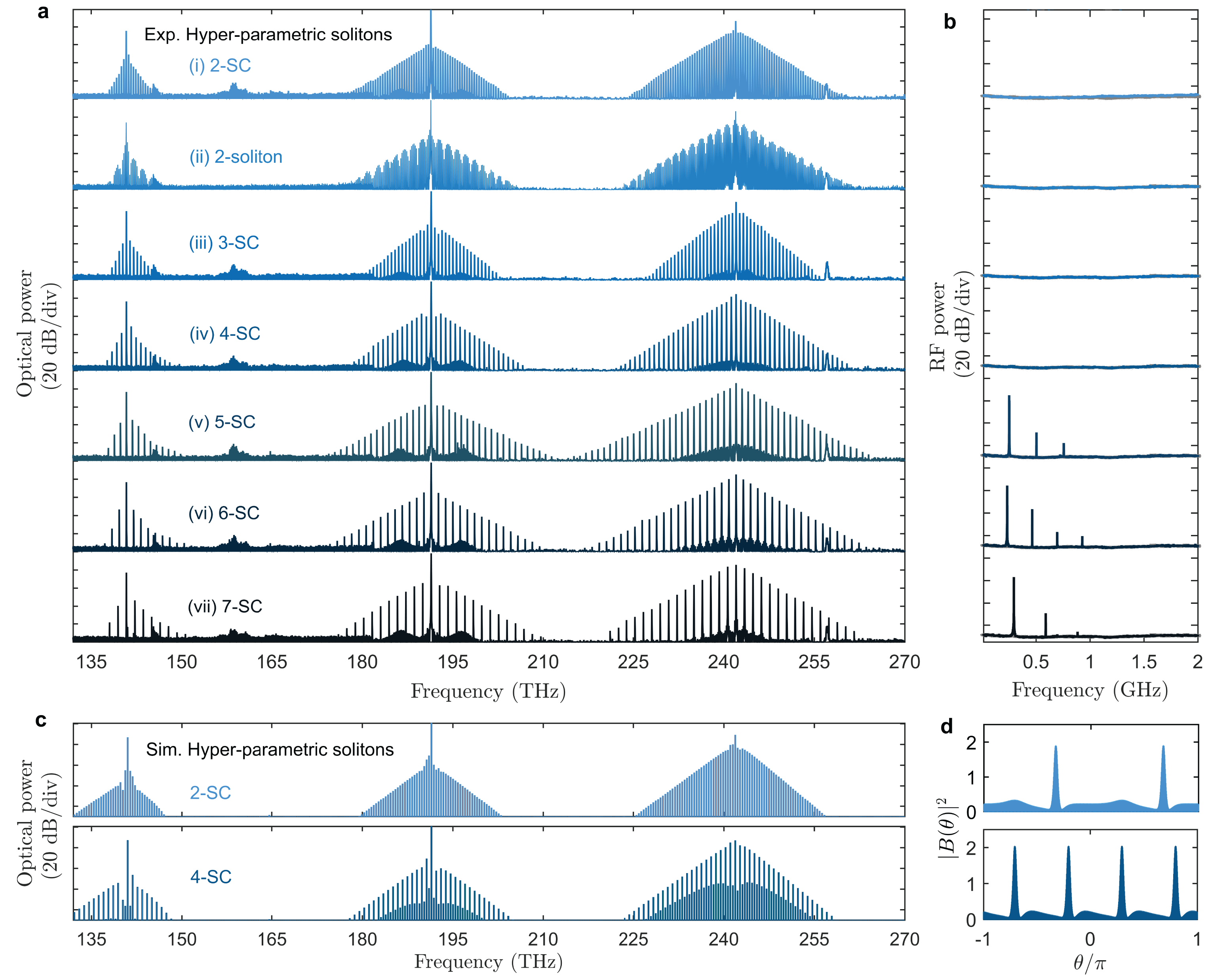}  
		\caption{{\bf Hyperparametric soliton crystals, quasi-crystals, multi-soliton states, and breathers.} {\bf a,} Experimentally measured three-colour spectra (top to bottom) for the single-soliton state, two-soliton state, three-soliton crystal, four-soliton crystal, and five-, six-, and seven-soliton-breather crystals. {\bf b,} RF spectra corresponding to the soliton states shown in (a). {\bf c,} Numerically modelled spectra of a two-soliton crystal and a four-soliton quasi-crystal. {\bf d,} Spatial profiles of signal pulses associated with the $240$THz spectra in (c).} 
		\label{f5}\end{figure*}

	\section*{Results}
	In this work, we used a Si$_3$N$_4$ microring resonator with a cross-section of 770 nm×2520 nm (thickness×ring width) and approximately 200GHz repetition rate designed to provide phase-matching conditions between two identical pump $TM_{00}$ photons with frequency around $191.4$THz and signal and idler photons close to $242.0$THz and 140.8THz, respectively. The details for designing the phase-matching and choosing the present geometry condition can be found in the Supplementary Information (Figs. S1a-S1c). The resonators were fabricated via a DUV-based subtractive approach, with the capability of wafer-scale manufacturing of ${{\rm Si}_3}{{\rm N}_4}$ photonic integrated circuits with ultra-low propagation loss and precise control of dimensions~\cite{hz1}, see Methods. A pulley coupler with a narrow gap of $650$nm was used to provide significant over coupling for the pump, see Figs. \ref{f2}a,b bringing the pump coupling rate above the one for the signal. A transmission spectrum shown in the Supplementary Information (Fig. S1d) confirms the significant over-coupling of the whole TM$_{00}$ mode family. Thus, the total, i.e., loaded,  quality factor, $Q^\mathrm{load}$, for the signal was above that for the pump, see Fig.  \ref{f2}c, which facilitated the comb generation to be triggered and dominated by the signal. The waveguide segment having the same curvature as the ring is approximately 50$\mu$m long, which yielded the anti-resonance coupling range around $154$THz  corresponding to the significant drop in the coupling rate and a sharp peak in  $Q^\mathrm{load}$~\cite{moille2}, see Fig.  \ref{f2}d. Available light sources allowed us to characterise losses around the pump while the whole range of $Q^\mathrm{load}$ values vs frequency was computed numerically, see Fig. \ref{f2}d.

	The resonant frequencies, $f_\mu$, and the four-wave mixing phase matching parameter, 
	\be
	\Delta f_\mu=f_{\mu}+f_{-\mu}-2f_0,~\mu=0,\pm 1,\pm 2,\dots,
	\lab{e1}\ee
	were computed numerically and measured for a group of modes around $190$THz, see Figs. \ref{f2}e and \ref{f2}f.  The signal and idler frequencies are determined from the  condition of the parametric gain achieving its threshold value which is most readily satisfied for a spread of modes in the proximity of the one best complying with $\Delta f_\mu=0$. Achieving $\Delta f_\mu=0$ without invoking nonlinear frequency shifts is possible for large values of $\mu$, when there is a change between anomalous and normal dispersion. A plot of the phase-matching parameter vs mode numbers, $\mu$, for our resonator is shown in Fig.~\ref{f2}f, see also prior results in Refs. \cite{murdoch,fujii,perez,kipp1,kartik11,gaeta1,victor,tang}. 
	Defining dispersion as $D_{2\mu}=f_{\mu+1}+f_{\mu-1}-2f_\mu$, we find that the groups of modes around the pump and signal have weakly anomalous dispersions, $D_{2p}= 1.7$MHz and $D_{2s}= 2.9$MHz,  while the idler dispersion  is large and normal, $D_{2i}=-22.7$MHz.

	The comb characterisation setup is schematically shown in Fig. \ref{f3}a, where the amplified pump is coupled into the microresonator using a mode-matched lensed fibre. For on-chip power of 100mW, we observed the excitation of monochromatic signal and idler waves with the frequencies predicted by zeros of the phase-matching parameter, see Fig. \ref{f3}b, broadly in agreement with prior work on Kerr OPOs~\cite{murdoch,fujii,perez,kipp1,kartik11,gaeta1,victor,tang}.
	We also tuned the signal and idler frequencies by tuning the pump to a range of modes, see Fig. S2 in Supplementary Information. 
	However, for the on-chip powers in the range of 370 to 500mW, we are reporting the hyperparametric soliton states corresponding to the three-colour low-noise frequency combs, see Figs. \ref{f3}c-\ref{f3}e, and 4, which we describe below in detail. The microresonator chip was placed on a piezo-positioning stage with a temperature controller to tune the phase-matching parameter. Solitons were reproducibly observed in different resonators while either applying or not applying temperature tuning.
	
	Figure \ref{f3}c shows the comb power traces as the pump wavelength increases (sweeping speed is 0.5 nm/s) for six fixed values of the on-chip laser power. The pump was suppressed and does not contribute to the power measurements in Fig. \ref{f3}c, where the green shading highlights the soliton steps. A representative soliton spectrum, shown in Fig.~\ref{f3}d, exhibits three triangle-shaped spectral bands centred around the idler, pump, and signal, with the signal band exhibiting the highest comb power by a significant margin. The comb repetition rate (line-to-line separation) was measured to be consistent across all three spectral bands, $199.5$GHz, and nearest to the linear repetition rate at the signal frequency, $199.75$GHz, see Fig. \ref{f3}e and Fig. \ref{f2}e. These observations indicate that all three comb components circulate within the resonator at the same angular rate locked by nonlinear effects, as expected for a three-component soliton, and that the signal wave plays a crucial role in the soliton generation process.

	Another pronounced feature of the soliton spectra is that the central signal mode at $\mu= 253$ dominates the comb spectrum generated around it. This is very much unlike what has been observed for the parametric solitons in degenerate OPOs~\cite{bruch,engl,moille}, which have a zero background and, therefore, their central frequency is blended with the rest of the comb spectrum, cf. Figs.~\ref{f1}d and \ref{f1}h. The strong central sideband in the signal comb in Fig. \ref{f3}d should be compared with no such feature in the parametric soliton data in Fig. 2a(iv) in \cite{bruch}, Fig. 4d in \cite{engl} and Figs. 4a, 4b in \cite{moille}. This effect is even more pronounced in the idler field, $\mu=-253$. Thus, we are dealing with the multi-colour modelocked pulses having the background shaped by the monochromatic parametrically generated signal and idler waves. 
	This observation prompted us to apply the term hyperparametric soliton as was already discussed in the previous section, see Fig.~\ref{f1}.
	
	Since no present concept or theory explains our measurements, it appears instructive to outline our numerical results and their interpretations. The first step in our analysis of mechanisms of the hyperparametric soliton generation was to solve the problem of three-wave parametric interaction of the $\mu=0$ mode with several pairs of the signal and idler in the proximity of  $\Delta f_\mu=0$ point, see Methods. Numerically computed signal components of all OPO states found above the parametric threshold for a chosen input power are shown in Fig.~\ref{f4}a. The pump, $\mu=0$ mode, becomes strongly depleted by the generated signal and idler pairs so that its dependence on the laser frequency deviates significantly from the parametric-process-free nonlinear resonance shown by the dashed line in Fig.~\ref{f4}b. 
	
	Notably, generating $|\mu|=253$, $254$, and $255$ states leads to the bistable OPOs. One should note that the negative-detuning range of the $253$ state stays isolated from the range of detunings where multiple other parametric states coexist,  see  Fig.~\ref{f4}a and Fig.~\ref{f1}f. We name the upper and lower branches of the bistability loop made by this state as OPO-$1$ and OPO-$2$. However, the instabilities of  OPO-$1$ and OPO-$2$ relative to the excitation of other modes (modulational instability) can not be excluded.
	
	To check stability and instability scenarios, we modelled a system of coupled equations for the amplitudes $a_\mu$ of the resonator modes $\mu=-512,\dots,0,\dots, 511$, covering the spectrum from $90$ to $290$THz, see Methods. We have found that the OPO-$1$ is typically modulationally unstable, and OPO-$2$ is typically stable through the detuning interval of their coexistence, which is the condition leading to the generation of hyperparametric solitons. The spectral carpet shown in Fig.~\ref{f4}d and computed by varying the laser frequency from large negative detunings, $\delta=(f_0-f_p)/\tfrac{1}{2}\kappa_0$, to positive ones, illustrates the modulational instability and soliton generation processes. Here, $f_p$ is the laser frequency and $\kappa_0$ is the width of pumped resonance. CW parametric generation begins at $\delta\approx-2.16$ and destabilises due to sideband generation at $\delta\approx -1.6$, forming three-colour frequency combs. These then evolve into hyperparametric solitons that exist for $\delta$ values in the $(-0.65,-0.06)$ range. 
	
	A numerically computed spectrum of the hyperparametric soliton (see Fig.~\ref{f4}c) qualitatively matches the typical experimentally observed spectra in Fig.~\ref{f3}d. The reconstructed pulse envelope profiles for a hyperparametric soliton's signal, pump and idler components are shown in Fig.~\ref{f4}e. It shows that the pump and idler waves are quasi-CW fields, and only the signal is a pronounced bright soliton resting on a background with finite amplitude. 
	
	The idler is not expected to make a pulse because its frequency lies in the interval of large normal dispersion. While the pump is in the anomalous dispersion range, it experiences a loss rate three times higher than the signal and also remains quasi-CW. For the pump to trigger soliton generation of its own, i.e., without the parametric process involved, requires the laser frequency to be tuned closer to the tip of its resonance ($\delta\approx 4$ on the dashed black line in Fig.~\ref{f4}b). 
	
	When spectra of the pump, signal and idler combs remain well isolated from each other, one can split the full coupled-mode system in equations for three modal groups and introduce the respective envelop functions, $A_p$, $A_s$, and $A_i$,  Fig.~\ref{f4} caption and Methods. By checking the signal equation, see Eq. \bref{tp16b} in Methods, reveals that the signal soliton centred on a mode $\mu'$ is excited by the quasi-CW nonlinear polarisation wave induced by the four-wave mixing process $A_p^2A_i^*\approx a_0^2a_{-\mu'}^*$. 
	Thus, the signal equation can be considered the generalised Lugiato-Lefever equation, where the pump has a complex dependence on the detuning introduced by the parametric process. The latter is anticipated to be the OPO-$2$ state. Diamonds, squares, and crosses in Fig.~\ref{f4}e show perfect matching between the hyperparametric soliton tails and the power values of the signal, pump and idler of the OPO-$2$ state, see Fig. \ref{f4}a. This finding unambiguously confirms that the hyperparametric soliton 
	have the background wave matching the OPO-2 state.
	
	By pumping different resonator modes, tuning pump power, and varying the phase matching parameter, we have further checked that conditions for the hyperparametric solitons to exist are that the first parametric state bifurcating from the pump for the negative detunings must become bistable and provide a range of parameters (pump power and detuning) where the lower (OPO-$2$) branch is stable and the upper (OPO-$1$) branch is modulationally unstable. These conditions are typically met when the net signal losses are sufficiently small and the power of the four-wave mixing term, $a_0^2a_{-\mu'}^*$, driving the signal mode $a_{\mu'}$, is sufficiently high. 
	
	For the soliton generation, we recorded a video, see Supplementary Information, showing how the signal and pump spectra evolve during the manual frequency tuning across the hyperparametric soliton range, see Supplementary Materials, where non-solitonic combs first replace monochromatic signal and pump, which then evolve to the soliton generation. The strong, $\approx 8$dBm, central sideband in the signal component rising above the comb starting around $-8$dBm, unambiguously points at the hyperparametric regime. The soliton existence on the negatively detuned tail of the resonance, where intraresonator fields and thermal effects are modest, enables the simplicity of the manual tuning. Tuning to the range of positive detunings boosts the circulating power, leading to the sudden loss of resonance due to thermal shifts. Experimental access to these regimes can be a subject of future work using one of the established techniques for observing positively detuned non-parametric solitons of the Lugiato-Lefever equation~\cite{kipgaeta,silver}. 
	
	Moreover, we demonstrated the robustness and extended parameter range of the hyperparametric solitons by varying the pumped modes and power,  see Fig. S3 in Supplementary Information. 
	The measured spectra can vary from a single soliton to soliton crystals or quasi-crystals, as shown in Fig.~\ref{f5}. A single hyperparametric soliton excites every mode, resulting in a triangular spectrum. In crystals containing \(N\) solitons, only every \(N\)th mode is excited, while quasi-crystals exhibit weaker lines between the strong spectral peaks. Additionally, multi-soliton non-crystal states, where \(N\) pulses are situated far apart from the \(2\pi/N\) separation distances, have also been frequently observed. RF noise measurements, presented in Fig. \ref{f5}b, confirm that the single, two-, three-, and four-soliton states are mostly stable, while five-, six-, and seven-soliton crystals are typically breathers \cite{ga,don}. 
	
	As the number of hyperparametric solitons in a crystal increases, the magnitude of the central sidebands in the idler and pump fields remains dominant. Conversely, the sidebands in the signal field decrease significantly. This is because the signal background rapidly loses power as the ring gets filled with pulses. Meanwhile, the pump and idler fields remain quasi-cw, with their power being only weakly dependent on the number of solitons in the crystal. This feature has been observed in both experimental and numerical data, as illustrated in Fig. \ref{f5}. 
	
	Soliton crystals have now emerged as a sub-field that connects photonics, condensed matter, and topological physics \cite{cole,kar,skrt,engt,mit,flt}, with the hyperparametric frequency conversion effect paving a new path for their investigation.

	\section*{Discussion}
	
	We demonstrated hyperparametric three-colour idler-pump-signal solitons in a nondegenerate Kerr microresonator OPO with a $200\mathrm{GHz}$ pulse repetition rate. The idler and pump soliton components exhibit quasi-CW behaviour, while the signal component forms a pronounced bright soliton pulse sitting on a parametrically generated background. This soliton regime arises when one of the above-threshold signal-idler pairs makes a hysteresis loop (bistability) from the modulationally unstable high-power and stable low-power OPO states. Our modelling shows that this OPO bistability occurs at negative pump detunings, which avoids the positive detuning region typically associated with pump-centered Lugiato–Lefever solitons. At negative detunings, intra-resonator powers remain relatively low, minimizing unwanted thermal effects. The quasi-CW nature of the pump and idler components is maintained through a combination of strong coupling losses at the pump (overcoupled resonator) and normal dispersion at the idler wavelength.

	Nondegenerate OPOs provide broad and flexible tunability through resonator geometry and temperature control of the refractive index and phase-matching conditions \cite{murdoch,fujii,perez,kipp1,kartik11,gaeta1,victor,tang}. This contrasts with degenerate OPOs, in which the signal frequency is strictly determined by the pump~\cite{bruch,engl,moille}. We also observed and reported the formation of crystals and breather states of hyperparametric solitons. Specifically, we generated solitons in the optical communication O-band using a C-band pump source.
	Expanding the availability of O-band light sources based on silicon photonics is a key milestone in the ongoing development of low-power optical interconnects for data center applications~\cite{bow}.
	
	Overall, our results on hyperparametric solitons open new avenues for studying their dynamics and interaction with other soliton types in nondegenerate Kerr and especially in $\chi^{(2)}$ microresonators, where realising the comb generation around widely separated and tunable signal and idler frequencies remains an open challenge.

	\section*{Methods}
	
	\noindent\textbf{Device fabrication and characterisation.} 
	The Si$_3$N$_4$ microresonators were fabricated using a deep ultraviolet (DUV) based subtractive process on 4-inch silicon wafers. A 4 $\mu$m thermal oxide layer was first grown as the bottom cladding.
	To mitigate the high tensile stress in thick Si$_3$N$_4$ films, interconnected stress-release trenches (3.5 $\mu$m depth) were patterned in the SiO$_2$ substrate using dry etching, spaced 10-30 $\mu$m from waveguide regions to ensure waveguide uniformity.
	A 770 nm thick stoichiometric Si$_3$N$_4$ layer was then deposited in a single low-pressure chemical vapour deposition (LPCVD) step, achieving 0.6$\%$ thickness uniformity across the wafer. 
	Waveguides were patterned using KrF 248 nm DUV stepper lithography with 180 nm resolution, followed by anisotropic dry etching with a C$_x$F$_y$ chemistry. An LPCVD SiO$_2$ hard mask was employed during etching to achieve smooth sidewalls with 87$^{\circ}$ sidewall angles. 
	The devices underwent high-temperature annealing at 1200°C for 11 hours in N$_2$ atmosphere to reduce hydrogen-related absorption losses. A 1.3 $\mu$m thick LPCVD SiO$_2$ top cladding was then deposited and subjected to identical annealing conditions. 
	
	\noindent\textbf{OFC measurements.} To characterize the resonator’s transmission, we swept a tunable laser across the 1480–1640 nm wavelength range with a high resolution of 0.1 pm and recorded the output optical power. This transmission spectrum enabled the extraction of key resonator parameters, including resonance linewidths, free spectral ranges (FSRs), and phase-matching conditions. We employed a tunable laser source amplified by an erbium-doped fibre amplifier (EDFA) coupled into the waveguide via a lensed fibre for hyperparametric soliton generation. Before the lensed fibre, a fibre polarization controller (FPC) was used to optimize the pump polarization, while a bandpass filter (BPF) was inserted to suppress amplified spontaneous emission (ASE) noise. The output light was collected using a second lensed fibre and split into two detection paths. One branch was analyzed with an optical spectrum analyzer (OSA) to monitor spectral features. At the same time, the other was routed through a fibre Bragg grating (FBG) for comb power monitoring and radio-frequency (RF) noise analysis using a digital storage oscilloscope (DSO) and an electrical spectrum analyzer (ESA). Access to optical parametric oscillation (OPO) sidebands and hyperparametric soliton states was achieved by gradually tuning the pump laser into resonance via continuous sweeping or manual adjustment. The typical soliton dynamics were captured in the Supplementary Video, which documents the manual tuning of the pump wavelength from 1566.37 nm to 1566.42 nm at an on-chip pump power of 425 mW.

	\noindent\textbf{Nonlinear modelling.}
	Multimode intra-resonator field is expressed  as
	\be
	\cA e^{iM\vta-i\om_p \tilde t}+c.c.,~\cA=\sum_\mu { a}_\mu(\tilde t) e^{i\mu\vta}.
	\lab{field}
	\ee
	Here,  $\vta=(0,2\pi]$ is the angular coordinate in the laboratory frame, $\tilde t$ is time and $a_\mu$ are mode amplitudes and $\mu=0,\pm 1,\pm 2,\dots$ is the relative mode number.     Coupled-mode equations governing nonlinear interactions between $a_\mu$ are ~\cite{chembo,herr,cont,prapl}
	\be
	\lab{tp9}
	\begin{split}
		&i\p_t a_{\mu}=-\gamma\sum_{\mu_1 \mu_2 
			\mu_3}\wh\delta_{\mu,\mu_1+\mu_2-\mu_3}a_{\mu_1}a_{\mu_2}a_{\mu_3}^*\\
		&+(f_{\mu }-f_p)a_{\mu} - \frac{\kappa_\mu}{2}
		\big(a_{\mu}-\wh\delta_{\mu,0}\sqrt{b\cW}\big).
	\end{split}
	\ee
	Here, $\wh\delta_{\mu,\mu'}=1$  for $\mu=\mu'$ and is zero otherwise.
	The resonator spectrum, $f_\mu=\om_\mu/2\pi$, and the mode number dependent linewidth, $\kappa_\mu$, were computed using Lumerical software, see Fig.~\ref{f2}. 
	Since we work with frequencies, $f_\mu$, and not the angular frequencies, $\om_\mu$, time $t$ in Eq.~\bref{tp9} was redefined as $t=2\pi \tilde t$.
	The pump laser frequency, $f_p$, is tuned around  $f_0=191.44$THz corresponding to the $M=820$ resonance. $\cW$ is the on-chip laser power, $b=179.5$ is the resonator power build-up factor for $f_p=f_0$.  $\gamma=f_0 n_2/S n_0$ is the nonlinear parameter~\cite{vahala}; $n_2=2.6\cdot 10^{-19}$m$^2$/W (Kerr coefficient), $S=1.32\cdot 10^{-12}$m$^2$ (mode area), $n_0=2.022$ (refractive index), $\gamma=18.65$MHz/W.  
	$|a_\mu|^2$ represent intraresonator powers of the comb lines scaled to have units of Watts~\cite{cont}.  Data shown in Figs. 
	\ref{f3}c-e, \ref{f4} are obtained by modelling Eq.~\bref{tp9}.
	While numerically solving Eq.~\bref{tp9}, we divided them by $\kappa_0/2$, $\kappa_0=683$MHz.  The scaling parameter applied  to plot dimensionless modal powers in Figs.~\ref{f4}a,b,e is $\kappa_0/2\gamma=18.3$W. Linewidth parameters are $\kappa_\mu=f_\mu/Q^\mathrm{load}_\mu$, see Fig.~\ref{f2}d and Table~\ref{tab1}.
	
	\begin{table}[t]
		\caption{\label{tab1}Normalised phase-matching  and linewidth parameters for $\pm\mu$ signal-idler pairs. }
		\begin{ruledtabular}
			\begin{tabular}{llllllll}
				$\mu$ & $250$ & $251$ & $252$ & $253$ & $254$ & $255$ & $256$ 
				\\ \hline
				$\Delta f_\mu/\kappa_0$ & $6.776$ & $5.442$ & $4.078$ & $2.683$ & $1.258$ & $-0.198$ & $-1.686$
				\\ \hline
				$\kappa_\mu/\kappa_0$ & $0.2785$ & $0.2760$ & $0.2734$ & $0.2709$ & $0.2685$ & $0.2660$ & $0.2636$ 
				\\ \hline\hline 
				$\mu$ & $-250$ & $-251$ & $-252$ & $-253$ & $-254$ & $-255$ & $-256$ 
				\\ \hline
				$\kappa_\mu/\kappa_0$ & $0.3281$ & $0.3388$ & $0.3496$ & $0.3607$ & $0.3719$ & $0.3833$ & $0.3949$
			\end{tabular}
		\end{ruledtabular}
		\lab{tab1}
	\end{table}
	
	Three-wave OPO states, represented by the pump mode and a signal-idler pair, see Figs. \ref{f4}a,b, were computed from the three-mode reduction of Eq.~\bref{tp9}.
	\be
	\lab{tp10}
	\begin{split}
		i\p_t a_0&=-2\gamma a_\mu a_{-\mu}a_0^*
		-\gamma(|a_0|^2+2|a_\mu|^2+2|a_{-\mu}|^2)a_0\\ &+(f_0-f_p) a_0 
		-i\frac{\kappa_0}{2}(a_0-\sqrt{b\cW}),
		\\
		i\p_t a_\mu&=
		-\gamma a_0^2a_{-\mu}^*-\gamma (|a_\mu|^2+2|a_0|^2+2|a_{-\mu}|^2)a_\mu\\ &+(f_\mu-f_p)a_\mu-i \frac{\kappa_\mu}{2} a_\mu,
		\\
		i\p_t a_{-\mu}&=-\gamma a_0^2a_{\mu}^*-
		\gamma (|a_{-\mu}|^2+2|a_0|^2+2|a_{\mu}|^2)a_{-\mu}\\ &+(f_{-\mu}-f_p)a_{-\mu}-i \frac{\kappa_{-\mu}}{2} a_{-\mu}.
	\end{split}
	\ee
	Parametric terms are placed first after the equal signs in all three equations. 
	The role of phase-matching parameter, $\Delta f_\mu$, see Eq.~\bref{e1}, in the three-wave model becomes explicit on the observation that
	\be
	f_{\pm\mu}-f_p=\frac{1}{2}\Delta f_\mu+(f_0-f_p)\pm\zeta_\mu,
	\lab{tp11}
	\ee
	while the $\zeta_\mu=(f_\mu-f_{-\mu})/2$ term  can be eliminated from Eq.~\bref{tp10} by substituting $a_{\pm\mu}=\wt a_{\pm\mu}\exp(\mp i\zeta_\mu t)$. 
	So that, Eq.~\bref{tp10} becomes
	\be
	\lab{tp12}
	\begin{split}
		i\p_t a_0&=-2\gamma \wt a_\mu \wt a_{-\mu}a_0^*
		-\gamma(|a_0|^2+2|\wt a_\mu|^2+2|\wt a_{-\mu}|^2)a_0\\ &+(f_0-f_p) a_0 
		-i\frac{\kappa_0}{2}(a_0-\sqrt{b\cW}),
		\\
		i\p_t \wt  a_\mu&=
		-\gamma a_0^2\wt a_{-\mu}^*-\gamma (|\wt a_\mu|^2+2|a_0|^2+2|\wt a_{-\mu}|^2)\wt a_\mu\\ &+(\tfrac{1}{2}\Delta f_\mu+f_0-f_p)a_\mu-i \frac{\kappa_\mu}{2} \wt a_\mu,
		\\
		i\p_t \wt a_{-\mu}&=-\gamma a_0^2\wt a_{\mu}^*-
		\gamma (|\wt a_{-\mu}|^2+2|a_0|^2+2|\wt a_{\mu}|^2)\wt a_{-\mu}\\ 
		&+(\tfrac{1}{2}\Delta f_\mu+f_{0}-f_p)\wt a_{-\mu}-i \frac{\kappa_{-\mu}}{2} \wt a_{-\mu}.
	\end{split}
	\ee
	Equation~\bref{tp12} explicitly shows that the OPO operation depends on the pump detuning, $f_0 - f_p$, and the phase-mismatch parameter, $\Delta f_\mu$. 
	
	Values of the relative linewidth parameters for the relevant signal-idler pairs are specified in Table~\ref{tab1}. Uncertainties in the parameter selection for numerical modelling come from the absence of experimental data on the loss values in the proximity of the signal and idler fields and temperature-induced fluctuations of the refractive index impacting phase-matching conditions. 
	
	Pulse envelopes, $A_p$, $A_s$ and $A_i$, corresponding to the well-separated pump, signal and idler spectra can be reconstructed from the modal amplitudes as 
	\be
	\lab{tp14}
	\begin{split}
		& A_p=\sum\nolimits_{\mu=-N}^{N}a_\mu e^{i\mu\theta},\\ 
		& A_s=e^{-i\zeta_{\mu_0}t}\tilde A_s,~\tilde A_s=\sum\nolimits_{\mu=\mu_0-N}^{\mu_0+N}\tilde a_\mu e^{i\mu\theta},\\  
		& A_i=e^{i\zeta_{\mu_0}t}\tilde A_i,~\tilde A_i=\sum\nolimits_{\mu=-\mu_0-N}^{-\mu_0+N}\tilde a_\mu e^{i\mu\theta},
	\end{split}
	\ee
	where $\mu_0$ and $N$ can be chosen, e.g., as $\mu_0=253$ and $N=100$.
	We have now replaced the angular variable $\vta$ with  $\ta=\vta-D_{1s}t$
	rotating with repetition rate at the signal frequency and will approximate dispersions only upto the second-order terms.  Equations for three envelopes, although their use for modelling comb dynamics is left for future work, allow for an understanding of the hyperparametric soliton generation at a qualitative level and contrast it with the previously reported parametric solitons~\cite{bruch,engl,moille}:
	\bsub
	\lab{tp16}
	\begin{align}
		i\p_t A_p&=-i(D_{1p}-D_{1s})\p_\ta A_p-\frac{1}{2}D_{2p}\p_\ta^2 A_p-2\gamma \wt A_s \wt A_i A_p^*
		\nn\\ 
		&+(\Delta_p-\gamma|A_p|^2) A_p -i\frac{\kappa_0}{2}(A_p-\sqrt{b\cW}),
		\label{tp16a}\\
		i\p_t \wt  A_s&=-\frac{1}{2}D_{2s}\p_\ta^2 \wt A_s-\gamma A_p^2\tilde A_i^*
		\nn\\ &+(\Delta_s-\gamma|\wt A_s|^2) \wt A_s -i\frac{\kappa_{\mu_0}}{2}\wt A_s,
		\label{tp16b}\\
		i\p_t \wt A_i&=-i(D_{1i}-D_{1s})\p_\ta \wt A_i-\frac{1}{2}D_{2i}\p_\ta^2 \wt A_i-\gamma A_p^2\tilde A_s^*
		\nn\\ 
		&+(\Delta_i-\gamma|\wt A_i|^2) \wt A_i -i\frac{\kappa_{-\mu_0}}{2}\wt A_i,	
		\label{tp16c}
	\end{align}
	\esub
	where we omitted higher-order dispersion, and introduced short-hand notations for effective detunings incorporating pump detuning, phase-mismatch, and nonlinear cross-phase modulation terms,
	\bsub
	\lab{tp15}
	\begin{align}
		&\Delta_p=f_0-f_p-2\gamma (|\wt A_s|^2+|\wt A_i|^2),
		\label{tp15a}\\
		&\Delta_s=\tfrac{1}{2}\Delta f_{\mu_0}+f_0-f_p-2\gamma (| A_p|^2+|\wt A_i|^2),
		\label{tp15b}\\
		&\Delta_i=\tfrac{1}{2}\Delta f_{\mu_0}+f_0-f_p-2\gamma (| A_p|^2+|\wt A_s|^2).
		\label{tp15c}
	\end{align}
	\esub
	
	Since, in the hyperparametric soliton regime, the pump and idler fields are quasi-CW, the $\Delta_s$ parameter can be approximately treated as a quasi-constant, $\p_\ta\Delta_s \approx 0$, and equations for $A_p(t,\ta)$ and $\wt A_i(t,\ta)$ can be approximated by equations for $a_0(t)$ and $\wt a_{-\mu_0}(t)$. Then, Eq.~\bref{tp16b} takes the form of the generalised Lugiato-Lefever model, where the four-wave mixing term, $A_p^2 \wt A_i^* \approx a_0^2 \wt a_{-\mu_0}^*$, $\p_\ta (A_p^2 \wt A_i^*)\approx 0$, plays the role of an effective pump, implicitly parametrised by the actual pump detuning, $f_0 - f_p$, and power, $\cW$. Here, $(D_{1p} - D_{1s})/\tfrac{1}{2}\kappa_0 = -2.2126$, $(D_{1i} - D_{1s})/\tfrac{1}{2}\kappa_0 = 2.8199$, $D_{2p}/\tfrac{1}{2}\kappa_0 =0.0052$, $D_{2s}/\tfrac{1}{2}\kappa_0 =0.0085$ and 
	$D_{2i}/\tfrac{1}{2}\kappa_0 =-0.0699$.
	
	Eqs.~\bref{tp16} are transformed to the equations for the degenerate Kerr OPO 
	used in Ref.~\cite{moille} to demonstrate parametric solitons, when $A_p$ is replaced with $A_s$, while $\wt A_s$ and $\wt A_i$ become two pump fields $A_{p1}$ and $A_{p2}$, so that the external pump term, $\sqrt{b\cW}$, is moved from Eq.~\bref{tp16a} to Eqs.~\bref{tp16b}, \bref{tp16c}. $A_{p1}$ and $A_{p2}$ can be approximated as quasi-CW
	fields~\cite{moille}, $\p_\ta (A_{p1}A_{p2}) \approx 0$, so that Eq.~\bref{tp16a} becomes a parametric Ginzburg-Landau equation,
	\bsub
	\lab{tp17}
	\begin{align}
		i\p_t A_s&=-\frac{1}{2}D_{2s}\p_\ta^2 A_s-2\gamma  A_{p1}  A_{p2} A_s^*
		\nn\\ 
		&+(\Delta_s-\gamma|A_s|^2 -i\frac{\kappa_0}{2})A_s,
		\label{tp17a}
	\end{align}
	\esub
	possessing explicit bright soliton solutions on zero background~\cite{miles,fl,barash,longhi,stal,engl,moille}.

	\vspace{8mm}\noindent{\bf Acknowledgements}\\
	This work was supported by a joint project between Research Ireland (SFI Grant No. 23/EPSRC/3920) and the UK Engineering and Physical Sciences Research Council (EPSRC Grant No. EP/X040844/1). Other support was received from  CONNECT Centre  (Grant No. 13/RC/2077.P2) and the Royal Society (Grant No. IES/R3/223225). 
	
	\vspace{3mm}\noindent{\bf Data availability}\\
	To request data supporting the findings inquire with D.V.S. and J.F.D.
	
	\vspace{3mm}\noindent{\bf Code availability}\\
	To request codes for data processing inquire with D.V.S.
	
	\vspace{1mm}\noindent{\bf Author contributions}\\
	H.W. performed the device design and measurements with the assistance from M.A., E.H.K., L.W., V.K., and W.G.. X.J. and T.J.K. fabricated the microresonators. D.V.S. developed theory and performed numerical simulations of OPO operation and solitons. H.W., J.F.D., and D.V.S wrote the manuscript text. D.V.S. and J.F.D conceptualised and supervised the project. All authors commented on the manuscript.
	
	\vspace{1mm}\noindent{\bf Competing interests}\\
	The authors declare no competing interests.
	
\end{document}